\documentclass{jmo}

\author{Jaromír KŘEPELKA}
\affil{Palacký University Olomouc, Faculty of Science, Joint Laboratory of Optics of Palacký University and Institute of Physics of the Czech Academy of Sciences, Olomouc, Czech Republic}

\authlist{J.~Křepelka}

\contacts{Jaromír Křepelka, Palacký University Olomouc, Faculty of Science, Joint Laboratory of Optics of Palacký University and Institute of Physics of the Czech Academy of Sciences, 17. listopadu 1192/12, 771 46 Olomouc, Czech Republic tel: +420 585 631 518, e-mail: jaromir.krepelka@upol.cz}

\title{Exact solution of maximally flat antireflection coatings for coherent and incoherent light}

\abstract{This paper presents two approaches to the precise design of maximally flat antireflection coatings reducing the reflectance of the substrate to near zero in a certain region around the central frequency. The first ideal case concerns coherent light interference, where it is required that for a chosen central frequency the maximum number of reflectance derivatives of the reflectance with respect to the frequency is zero. This approach makes it possible to determine the refractive indices of a system of homogeneous quarter-wave thin films by solving set of nonlinear equations that can be found in explicit form for a maximum of four layers; for higher numbers of layers, the result must be sought numerically. The second case is the incoherent superposition of light waves, which refers to the determination of the refractive indices of the layers independently of their thicknesses. For these two physically ideal cases, the obtained refractive indices of the layers and the dependence of their reflectivity on the light frequency are compared. A simple new method for approximating the refractive indices of a maximally flat system of antireflective layers is also proposed.}

\keywords{thin layers, thick layers, maximally flat antireflection coating, exact solution, coherent light, incoherent light}

\anntitle{Přesné řešení maximálně plochých antireflexních vrstev pro koherentní a~nekoherentní světlo}

\annabstract{Tento článek předkládá dva přístupy k přesnému řešení maximálně plochých antireflexních vrstev snižujících odraznost substrátu téměř na nulu v určité oblasti kolem centrální frekvence. První ideální případ se týká koherentní interference světla, kde je požadováno, aby pro zvolenou centrální frekvenci byl co nejvyšší počet derivací odraznosti podle frekvence nulový. Tento přístup umožňuje určit indexy lomu soustavy homogenních čtvrtvlnných tenkých vrstev řešením nelineárních rovnic, které lze nalézt v explicitní formě maximálně pro čtyři vrstvy, pro vyšší počet vrstev je nutné výsledek hledat numericky. Druhým případem je nekoherentní superpozice světelných vln, což se týká stanovení indexů lomu vrstev nezávisle na jejich tloušťce. Pro tyto dva fyzikálně ideální případy jsou porovnány získané indexy lomu vrstev a závislost jejich odraznosti na frekvenci světla. Rovněž je navržena jednoduchá nová metoda pro aproximaci indexů lomu maximálně plochého systému antireflexních vrstev.}

\annkeywords{tenké vrstvy, tlusté vrstvy, maximálně plochá antireflexe, přesné řešení, koherentní světlo, nekoherentní světlo}

\acknowledgement{The author (orcid.org/0000-0003-0684-0775) thanks project OP PIK No. CZ.01.1.02/0.0/0.0/21\_374/0027282 ``Relief nano/micro structures for optical components in the automotive industry'' SPP 843102011 and LM2023032 ``Pierre Auger Observatory" of the Ministry of Education, Youth and Sports of the Czech Republic for their support.}

\setboolean{researchpaper}{true}

\begin{document}
% Vlozeni titulku (jmena, afiliace, nazev, abstrakt a~klicova slova cesky)
\maketitle

% Textova sekce do dvou sloupcu
\begin{multicols}{2}

\section{Introduction}
Antireflection coatings are widely used in optical applications to reduce the reflectance of optical surfaces, e.g. for lenses applied as components of various devices, photovoltaic cells, optical sensors and more. Their purpose is to increase the light transmittance of optical system, with the side effect of improving image contrast by removing stray light caused by multiple internal reflections. This is especially important for the perfect function of photographic cameras, video cameras, binoculars, telescopes, laser devices, etc. To ensure such function, systems of layers are essential, the mathematically motivated improvement of which began with Strong's \cite{Strong1936} analysis of the reflectance of a~substrate coated with a~single layer of arbitrary thickness based on Fresnel's relations, which led to the frequently mentioned condition that the magnitude of the refractive index of the layer should be equal to the geometric average of the refractive indices of the substrate and the superstrate materials. This was followed by further theoretical advances in the analysis of the reflection properties of more complex systems, in particular thin interference layers, based on the principle of coherent summation of partial reflections of light waves from each material boundary. Among many researchers, we should mention Antonín Vašíček \cite{Vasicek1956}, a~Czechoslovak pioneer in the theory, preparation and measurement of optical systems with thin or thick layers. Mathematical simplification of seemingly somewhat confusing and complex calculations of macroscopically measurable parameters of thin film systems, based on the transformation of tangential components of electric and magnetic field intensities of plane monofrequency electromagnetic waves in isotropic homogeneous media, was then achieved by a~large number of authors starting with F. Abèles \cite{Abeles1948}. Let us remember, for example, the Czechoslovak thin film expert Zdeněk Knittl \cite{Knittl1976}. The matrix approach to the calculation of parameters of thin film systems for plane waves, representing the transformation of tangential components of plane electromagnetic waves, was generalized by Dwight W. Berremann for anisotropic media \cite{Berremann1967}.

The theoretical research possibilities have been extended, among other things, finding such layer systems that would reduce the reflectance of the substrate as much as possible over a~very wide range of light spectra. For instance in \cite{Krepelka1992} an exact solution for the refractive indices of quarter-wave systems of isotropic homogeneous dielectric layers is given for such an impedance matching, which, assuming perfect interference of all internally reflected waves (that is for coherent light), achieve maximally flat reflectance versus frequency (wavelength) in the sense that the maximum possible number of derivatives of reflectance versus frequency (or wavelength) for a~chosen central frequency (central wavelength) is equal to zero. Such a~mathematically rigorous problem leads to finding the roots of a~system of nonlinear equations, which has an explicit solution for a~maximum of 4 layers, for a~larger number of layers it is necessary to find a~solution using numerical methods. In this paper, this approach valid for ideal thin layers is compared with the approach applicable to ideally thick layers, where none of the internally reflected waves interferes with the other. Such an aim mathematically requires finding the minimum of a~function of several variables. It should be noted that the obtained refractive indices of the layers in both cases require their gradual change from the refractive index of the superstrate towards the refractive index of the substrate (or vice versa), which are usually refractive indices that are not readily available technologically. However, although theoretical research on the design of antireflection layers seems to have been completed, with the development of nanotechnology, new possibilities for the realization of antireflection surfaces are beginning to emerge (see e.g. \cite{Schulz2023}).

\section{Theory}
\subsection{Reflectance and transmittance of thin and thick layers}
Let us consider the propagation of a~plane monofrequency electromagnetic wave with angular frequency $\omega$ (angular wave number in vacuum is $\omega /c$, $c$ is the speed of light in vacuum) in the geometry of parallel plane layers in a~stratified isotropic medium \cite{Krepelka1991}. There are $k$ homogeneous layers deposited on the substrate (rear half-space), the front half-space is superstrate. Let the electromagnetic wave incident the system from the superstrate be denoted as R (right) and from the opposite direction as L (left). Plane waves carry the same R or L designation even inside thin layers, the resulting superposed electromagnetic field is then a~linear combination of both counter propagating waves. Let us number the layers in the R direction of the incident wave, starting with 0 (for the superstrate), 1 (the layer last deposited on the substrate), 2, \dots, $k$ (the layer deposited on the substrate as the first), $k+1$ (the substrate). Each medium is characterized by its generally complex refractive index $n_j$, $j=0, 1, \dots , k+1$, the imaginary part of which represents an absorption due to the conductance of the materials, for ideal dielectric environments the refractive indices are real quantities. 

For simplicity, assume that the superstrate is a~dielectric (e.g. air) in which the direction of propagation of the incident R wave is determined by the angle of incidence $\theta_0$ measured from the normal of the plane material interface surfaces. Since the law of refraction $n_j\sin(\theta_j)=n_0\sin(\theta_0)$ holds, i.e. the product of the refractive index and the sinus of the propagation angle is invariant for all envolved media (we do not discuss the physical meaning of the generally complex propagation angles $\theta_j$), the normal component of the propagation vector in the $j$-th medium is
\begin{equation*}
\pm\dfrac{\omega}{c}\sqrt{n_j^2-(n_0\sin\theta_0)^2}=\pm\dfrac{\omega}{c}n_j\cos(\theta_j). 
\end{equation*}
Here, the plus sign applies to the L wave and the minus sign to the R wave in order to preserve the sense of the direction of wave propagation, assuming that the time dependence of the plane wave is a~harmonic function of time $t$ as $\exp(\mathrm {i}\omega t)$ and the imaginary parts of the refractive indices of the absorbing materials are chosen to be nonpositive. Maxwell's equations provide a~solution for such a~configuration, which can be briefly described in the terms below. Denoting $Z_0$ the vacuum impedance, then the admittance (ratio of tangential components of magnetic and electric field intensities) of the $j$-th medium is
\begin{equation}
Y_j =\frac{1}{Z_0}
\times 
\left\{
\begin{matrix}
\sqrt{n_j^2-(n_0\sin\theta_0)^2}=n_j\cos(\theta_j), \\
\dfrac{n_j^2}{\sqrt{n_j^2-(n_0\sin\theta_0)^2}}
=
\dfrac{n_j}{\cos(\theta_j)}, \\
\end{matrix}
\right.
\label{eq1}
\end{equation}
where the upper row defines the admittance of the $j$-th medium for the s (TE) wave and the lower row for the p (TM) wave. The correct calculation of square roots in the complex number domain should be taken into account when focusing on absorbing materials or the case of total internal reflection in dielectric materials.

Vectors of tangential components of the electric $\mathbf{E}_{\mathrm{t},j}$ and magnetic $\mathbf{H}_{\mathrm{t},j}$ field intensities composed of the counter propagating R ($\mathbf{E}_{\mathrm{tR},j}$, $\mathbf{H}_{\mathrm{tR},j}$) and L ($\mathbf{E}_{\mathrm{tL} ,j}$, $\mathbf{H}_{\mathrm{tL},j}$) waves in the $j$-th environment, are obtained from the tangential components of (only) electric field intensity but decomposed into R and L ($\mathbf{E}_{\mathrm{tR},j}$, $\mathbf{E}_{\mathrm{tL},j}$) waves using the transformation matrix
\begin{equation}
\begin{pmatrix}
\mathbf{E}_{\mathrm{t},j}\\
\mathbf{H}_{\mathrm{t},j}\\
\end{pmatrix}
=
\begin{pmatrix}
 \pm 1 & 1 \\
 Y_j & \mp Y_j\\
\end{pmatrix}
\begin{pmatrix}
\mathbf{E}_{\mathrm{tR},j}\\
\mathbf{E}_{\mathrm{tL},j}\\
\end{pmatrix},
\label{eq2}
\end{equation}
where the upper sign applies to the s-polarized waves and the lower sign to the p-polarized waves.

We conclude that the signs following directly from Maxwell's equations (and often omitted in calculations) are necessary for the correct determination of the phases of the amplitude reflectances of s and p waves, as it is known from Fresnel's formulas for a~single boundary between two materials, otherwise the p wave argument of amplitude reflectance would differ by $\pi$ from the correct value, which may affect, for example, the evaluation of ellipsometric measurements or modal analysis of planar waveguides. It should be noted that when calculating (power) reflectances as quadratic quantities, the change of sign for the p wave does not apply.

The tangential components of the electric field intensity, but decomposed into R and L waves, are obtained from the tangential components of the composed electric and magnetic field intensities using the inverse of (\ref{eq2})
\begin{equation}
\begin{pmatrix}
\mathbf{E}_{\mathrm{tR},j}\\
\mathbf{E}_{\mathrm{tL},j}\\
\end{pmatrix}
=
\frac{1}{2}
\begin{pmatrix}
 \pm 1 & \dfrac{1}{Y_j}\\
 1 & \mp \dfrac{1}{Y_j}\\
\end{pmatrix}
\begin{pmatrix}
\mathbf{E}_{\mathrm{t},j}\\
\mathbf{H}_{\mathrm{t},j}\\
\end{pmatrix}.
\label{eq3}
\end{equation}

If we introduce for the phase change of the plane wave as its propagates between the boundaries of the $j$-th layer of thickness $h_j$ the term $\varphi_j=(\omega/c)h_j\sqrt{n_j^2-(n_0\sin\theta_0)^2}$, then for the transformation of the tangential components of the electric and magnetic (composed) field intensities between the boundaries of the $j$-th layer to the tangential components of the electric field intensity of the counter propagating waves, we obtain for each of the perpendicular polarizations the interference matrix $\mathbf{M}_j$
\begin{equation}\label{eq4}
\begin{split}
&\mathbf{M}_j = 
\begin{pmatrix}
\cos(\varphi_j) & \pm \dfrac{\mathrm{i}}{Y_j}\sin(\varphi_j) \\
\pm {\mathrm{i}}Y_j\sin(\varphi_j) & \cos(\varphi_j)
\end{pmatrix},\\
&\begin{pmatrix}
\mathbf{E}_{\mathrm{t},j}\\
\mathbf{H}_{\mathrm{t},j}\\
\end{pmatrix}
=
\mathbf{M}_j
\begin{pmatrix}
\mathbf{E}_{\mathrm{t},j+1}\\
\mathbf{H}_{\mathrm{t},j+1}\\
\end{pmatrix}.
\end{split}
\end{equation}

Note that the (algebraic) spectral decomposition of the interference matrix $\mathbf{M}_j$ contains its eigenvectors identical to the columns of the matrix defined in equation (\ref{eq2})
\begin{equation}\label{eq5}
\begin{split}
&\mathbf{M}_j =
\begin{pmatrix}
 \pm 1 & 1 \\
 Y_j & \mp Y_j\\
\end{pmatrix}
\begin{pmatrix}
 \exp(\mathrm{i}\varphi_j) & 0 \\
 0 & \exp(-\mathrm{i}\varphi_j) \\
\end{pmatrix}\\
&\times\frac{1}{2}
\begin{pmatrix}
 \pm 1 & \dfrac{1}{Y_j}\\
 1 & \mp \dfrac{1}{Y_j}\\
\end{pmatrix}.
\end{split}
\end{equation}
In some cases, it is advantageous to use the decomposition of interference matrices in their eigenvectors and eigenvalues for an alternative calculation of the amplitude parameters of thin layers, in which explicit expressions for Fresnel reflectances (transmittances) at the boundaries of adjacent materials occur. From this decomposition it is clear that the matrix approach to the calculation of the thin film parameters is equivalent to an infinite sum of partially reflected and transmitted waves with phases shifted by $\varphi_j=\omega\tau_j=2\pi\nu\tau_j$ during each of their passage through the layers. Such an approach, with use of the normalized spectral power density of the source radiation and the Wiener-Khinchin theorem, also allows us to show (e.g., \cite{Bajer1991}) how the smooth transition from coherent to incoherent light works in stratified media.

Then the resulting matrix transmitting the tangential components of the electric field intensity decomposed into counter propagating waves between the outer boundaries of the system of $k$ layers is obtained by multiplying the interference matrices in the correct order
\begin{equation}
\mathbf{S}= 
\frac{1}{2}
\begin{pmatrix}
 \pm 1 & \dfrac{1}{Y_0}\\
 1 & \mp \dfrac{1}{Y_0}\\
\end{pmatrix}
\mathbf{M}_1\mathbf{M}_2\dots\mathbf{M}_k
\begin{pmatrix}
 \pm 1 & 1 \\
 Y_{k+1} & \mp Y_{k+1}\\
\end{pmatrix},
\label{eq6}
\end{equation}

\begin{equation}
\begin{pmatrix}
\mathbf{E}_{\mathrm{tR},0}\\
\mathbf{E}_{\mathrm{tL},0}\\
\end{pmatrix}
=
\mathbf{S}
\begin{pmatrix}
\mathbf{E}_{\mathrm{tR},k+1}\\
\mathbf{E}_{\mathrm{tL},k+1}\\
\end{pmatrix}.
\label{eq7}
\end{equation}
The relation (\ref{eq6}) for the transfer matrix $\mathbf{S}$ holds if the tangential components of the electric and magnetic field intensities are continuous at the boundaries of the layers, which is not generally fullfiled when surface currents or surface charges occur at the boundaries between the layers.

Using the standard definition of amplitude reflectances $r$ and transmittances $t$ of waves incidenting the layer system from the superstrate (subscript R) and from the substrate (subscript L) directions, we can calculate these macroscopic complex quantities from the elements of the transfer matrix $\mathbf{S}$ as follows
\begin{equation}
r_{\mathrm{R}}=\frac{S_{21}}{S_{11}},\ t_{\mathrm{R}}=\frac{1}{S_{11}},\ 
r_{\mathrm{L}}=-\frac{S_{12}}{S_{11}},\ t_{\mathrm{L}}=\frac{Y_{k+1}}{Y_0}\frac{1}{S_{11}},
\label{eq8}
\end{equation}
from where we immediately have
\begin{equation}
\mathbf{S} = 
\frac{1}{t_{\mathrm{R}}}
\begin{pmatrix}
1 & -r_{\mathrm{L}} \\
r_{\mathrm{R}} & d
\end{pmatrix},\ 
d=t_{\mathrm{R}}t_{\mathrm{L}}-r_{\mathrm{R}}r_{\mathrm{L}}.
\label{eq9}
\end{equation}

From the relation between inverse interference matrices and interference matrices as functions of $\varphi_j$ or $n_j$ for the cases of dielectric or absorbing media and from the physical requirement imposed on the transfer matrix $\mathbf{S}$ defined by equation (\ref{eq6}) for the same layer system in mirror symmetry, we can derive the correlation between the elements of the transmission matrix and hence the relation for the amplitude reflectances and transmittances from left and right (see \cite{Krepelka1983}). Using this so-called theorem of reversibility, we obtain for non-absorbing systems $S_{22}=S_{11}^*$, $S_{21}=S_{12}^*$), i.e. that the amplitude reflectances and transmittances of the system of thin dielectric layers (4 complex numbers) are definitely determined by only three real numbers using also the condition $\mathrm{det}(\mathbf{S})=Y_{k+1}/Y_0$. For instance we can choose for this purpose the absolute value of the amplitude reflectance from the left and its argument and the argument of the amplitude transmittance from the left and express with their help all macroscopic parameters. This approach based on the properties of the transfer matrix with respect to mirror symmetry was generalized for systems of anisotropic thin layers in \cite{Krepelka1993}.

Due to the dependence of the divergence of the energy flux of a~plane electromagnetic wave only on the normal (perpendicular) Cartesian coordinate, the power (energy) transfer in a~system of thin layers in a~planar arrangement is determined only by normal component of the Poynting vector calculated using tangential vectors of the intensity of the electric and magnetic fields separately for each of the counter propagating waves. Therefore, we will calculate the (power) reflectances and transmittances of the thin layer system following the relations

\begin{equation}\label{eq10}
\begin{split}
&\rho_{\mathrm{R}}=\left|r_{\mathrm{R}}\right|^2,\
\tau_{\mathrm{R}}=\frac{\mathrm{Re}(Y_{k+1})}{\mathrm{Re}(Y_{0})}\left|t_{\mathrm{R}}\right|^2,\\
&\rho_{\mathrm{L}}=\left|r_{\mathrm{L}}\right|^2,
\tau_{\mathrm{L}}=\frac{\mathrm{Re}(Y_{0})}{\mathrm{Re}(Y_{k+1})}\left|t_{\mathrm{L}}\right|^2,
\end{split}
\end{equation}
where Re stands for the real part of a~complex number. From this definition we obtain the transfer matrix for the normal components of Poynting vector
\begin{equation}
\mathbf{N} = 
\frac{1}{\tau_{\mathrm{R}}}
\begin{pmatrix}
1 & -\rho_{\mathrm{L}} \\
\rho_{\mathrm R} & \delta
\end{pmatrix},\ 
\delta=\tau_{\mathrm{R}}\tau_{\mathrm{L}}-\rho_{\mathrm{R}}\rho_{\mathrm{L}}.
\label{eq11}
\end{equation}
It means that for the normal components of Poynting vector at both sides of the thin film system we have 
\begin{equation}
\begin{pmatrix}
P_{\mathrm{nR},0} \\
P_{\mathrm{nL},0}
\end{pmatrix}
=\mathbf{N}
\begin{pmatrix}
P_{\mathrm{nR},{k+1}} \\
P_{\mathrm{nL},{k+1}}
\end{pmatrix}.
\label{eq12}
\end{equation}
%In this way, the parameters of a~system of thin layers in coherent light can be calculated.

If we are interested in the transformation of normal components of the Poynting vector of the waves propagating in one direction or the other, and thus in determining the reflectance and transmittance of a~thick layer, we must first determine the attenuation of the field propagating through a~layer of thickness $h$ with complex refractive index $n$, if the propagation angle $\theta_0$ is measured in a~medium with refractive index $n_0$. This attenuation of the field is determined by the attenuation factor $0 < U \le 1$
\begin{equation}
U=\exp\left(-2h\frac{\omega}{c}\left|\mathrm{Im}\sqrt{n^2-(n_0\theta_0)^2} \right|\right),
\label{eq13}
\end{equation}
 where the absolute value in the argument of the exp function is given to emphasize the energy passivity of the medium. With its help we can determine the transfer matrix of the normal components of the Poynting vector inside arbitrary thick layer
\begin{equation}
\mathbf{N}=
\begin{pmatrix}
1/U & 0 \\
0&U
\end{pmatrix}.
\label{eq14}
\end{equation}

The relations for the transfer matrices of the normal components of the Poynting vector allow them to be combined for any number of systems alternating thick layers with systems of thin layers (some of which may be empty), just multiplying them in the correct order. This is of course true if the tangential components of the electric and magnetic field intensities, and hence the normal components of the Poynting vector, are continuous at the boundaries. The resulting product of these matrices then needs only to be compared with the matrix elements in the equation (\ref{eq11}) to obtain the resulting (power) reflectances and transmittances. For example, for two thin film systems deposited on both sides of a~thick substrate, the resulting reflectances and transmittances are determined from the relation
\begin{equation}\label{eq15}
\begin{split}
&\frac{1}{\tau_{\mathrm{R}}}
\begin{pmatrix}
1 & -\rho_{\mathrm{L}} \\
\rho_{\mathrm{R}} & \delta
\end{pmatrix}
=
\frac{1}{\tau_{{\mathrm{R}}1}}
\begin{pmatrix}
1 & -\rho_{{\mathrm{L}}1} \\
\rho_{{\mathrm{R}}1} & \delta_1
\end{pmatrix}
\begin{pmatrix}
1/U_2 & 0 \\
0&U_2
\end{pmatrix}\\
&\times
\frac{1}{\tau_{{\mathrm R}3}}
\begin{pmatrix}
1 & -\rho_{{\mathrm L}3} \\
\rho_{{\mathrm R}3} & \delta_3
\end{pmatrix}
\end{split}
\end{equation}
where $\delta$ has the same meaning for the whole system as in (\ref{eq11}) and the quantities denoted here by the subscript 1 refer to a~system of thin layers deposited from the left on a~thick substrate and calculated from the equation (\ref{eq8}) assuming an infinitely extensive substrate from the right. The attenuation factor $U_2$ (equal to one for a~dielectric medium) refers to the thick layer material, and the quantities denoted by the subscript 3 refer to the system of thin layers deposited on the substrate from the right, while the thick layer material serves as an half-infinite medium on the left. Any of the subsystems may be empty.

In this way, for example, the influence of the reflection from the rear boundary of the substrate on the measured parameters of the layers can be taken into account, which is otherwise considered as a~half-space from which light is not reflected back to the system.

\subsection{Maximally flat antireflection system of layers for coherent light}
The assumption of perfectly coherent light is equivalent to the propagation of a~monofrequency field in ideal thin dielectric layers, whose interference effect is highest when the optical thicknesses of the layers are equal to a~quarter of the chosen (central) wavelength or integer multiples thereof. Therefore, let us look for the refractive indices of the system of quarter-wave layers that best minimize the reflectance of the substrate around the central wavelength at perpendicular incidence.

Let us define maximally flat antireflection coatings as those that exhibit zero value of the highest possible number of reflectance derivatives with respect to the angular frequency $\omega$ for the central frequency $\omega_c$. Derivatives can alternatively be made with respect to the common phase change as the wave pass through each layer, i.e., the common phase layer thickness $\varphi=\varphi_j = (\pi/2)\omega/\omega_c$, $j=1, \dots, k$. At the same time, the zero of the first derivative follows from the requirement of zero reflectance at the central wavelength, as it is a~local minimum of reflectance.

From the Taylor expansion of the interference matrices around $\varphi =\pi/2$ and the above requirement that the derivatives must be zero, we obtain a~system of nonlinear equations of the form \cite{Krepelka1992}

\begin{equation}
\begin{matrix}
N_{k_j}^{(a)}=n_0 n_{k+1}N_{k_j}^{(b)}, &1 \leq j \leq k, &j \ \mathrm{odd}, \\
N_{k_j}^{(a)}=\dfrac {n_0} {n_{k+1}}N_{k_j}^{(b)}, &2 \leq j \leq k, & j \ \mathrm{even},
\end{matrix}
\label{eq16}
\end{equation}
where the Pohlack's coefficients are expressed by the formulas \cite{Pohlack1952}

\begin{equation}
\begin{matrix}
N^{(a)}_{k_{j}}=\displaystyle\sum_{M_{k_{j}}}
\dfrac{n_{i_{1}}n_{i_{3}}\dots n_{i_{j}}}
{n_{i_{2}}n_{i_{4}}\dots n_{i_{j-1}}},
&1 \leq j \leq k, &j \ \mathrm{odd}, \\
N^{(a)}_{k_{j}}=\displaystyle\sum_{M_{k_{j}}}
\dfrac
{n_{i_{1}}n_{i_{3}}\dots n_{i_{j-1}}}
{n_{i_{2}}n_{i_{4}}\dots n_{i_{j}}}
&2 \leq j \leq k, & j \ \mathrm{even}, \\
N^{(b)}_{k_{j}}=\displaystyle\sum_{M_{k_{j}}}
\dfrac{n_{i_{2}}n_{i_{4}}\dots n_{i_{j-1}}}
{n_{i_{1}}n_{i_{3}}\dots n_{i_{j}}},
&1 \leq j \leq k, &j \ \mathrm{odd}, \\
N^{(b)}_{k_{j}}=\displaystyle\sum_{M_{k_{j}}}
\dfrac
{n_{i_{2}}n_{i_{4}}\dots n_{i_{j}}}
{n_{i_{1}}n_{i_{3}}\dots n_{i_{j-1}}}
&2 \leq j \leq k, & j \ \mathrm{even}.
\end{matrix}
\label{eq17}
\end{equation}
The summation set $M_{k_{j}}=\left\{ i_{1}, i_{2}, \dots i_{j} \right\}$ is determined by the inequalities
\begin{equation}\label{eq18}
\begin{split}
&1\leq i_{1}\leq k-(j-1),\ i_{1}+1\leq i_{2}\leq k-(j-2), \dots ,\\
&i_{j-1}+1\leq i_{j}\leq k-(j-j)
\end{split}
\end{equation}
or equivalently
\begin{equation}\label{eq19}
\begin{split}
&j\leq i_{j}\leq k,\ j-1\leq i_{j-1}\leq i_{j}-1, \dots,\\
&j-(j-1)\leq i_{1}\leq i_{2}-1.
\end{split}
\end{equation}

If both conditions (\ref{eq16}) are satisfied, then the reflectance of the maximally flat antireflection system defined in this way is depending on the common phase layer thickness $\varphi$ as follows

\begin{equation}
\rho(\varphi )=\dfrac{\rho_{0} \cos^{2k}\varphi}{1-\rho_{0}+\rho_{0}\cos^{2k}\varphi},
\label{eq20}
\end{equation}
where $\rho_{0}=\left[({n_{0}-n_{k+1})/(n_{0}+n_{k+1}})\right]^{2}$ is the reflectance of the bare substrate without layers that a~maximally flat antireflection system periodically achieves at frequencies equal to even multiples of the central frequency.

From the equivalence of summation sets (\ref{eq18}) and (\ref{eq19}) and the transformation of the indices $i_{j+1-m}\to i_{m},\ m=1, 2,\dots, j$ we obtain partial result that the first of the equations (\ref{eq16}) is solved if following symmetry condition between the refractive indices of the layers is satisfied

\begin{equation}
n_{j}n_{k+1-j}=n_{0} n_{k+1},\quad j=1, 2, \dots , k.
\label{eq21}
\end{equation}

The second set of equations (\ref{eq16}) together with the conditions (\ref{eq21}) represent a~complete system of nonlinear equations for unknown refractive indices. In \cite{Krepelka1992} it is shown that for one to four layers the solution (\ref{eq16}) can be found in explicit form, for a~higher number of layers it is necessary to find the solution of the system of equations of several variables numerically. The time required for the numerical calculation time increases dramatically with the number of layers, and the accuracy of the result is affected by the finite number of digits implemented in the computer software.

\subsection{Maximally flat antireflection system of layers for incoherent light}
The reflectance of a~system of $k$ thick dielectric layers with refractive indices $n_j$, $j=1,\dots, k$ located between media with refractive indices $n_0$ (for example air) and $n_{k+1}=n_s$ (for example substrate) is determined analogously according to the relation (\ref{eq15}) by multiplying the transfer matrices of the normal components of the Poynting vector at each boundary, where the matrices with attenuation coefficients are unit. In more detail, for zero angle of incidence, the reflectance of each interface from the left is given by the Fresnel relation

\begin{equation}
\rho_{\mathrm{R},j}=\rho_{j}=\left(\dfrac{n_{j-1}-n_j}{n_{j-1}+n_j}\right)^2,\quad j=1, 2, \dots , k+1.
\label{eq22}
\end{equation}

The reflectance from the right is the same as from the left $\rho_{\mathrm{L},j}=\rho_{\mathrm{R},j}$, transmittances are $\tau_{\mathrm{R},j}=\tau_{\mathrm{L},j}=1-\rho_j$ and the parameter $\delta_j=\tau_{\mathrm{R},j}\tau_{\mathrm{L},j}-\rho_{\mathrm{R},j}\rho_{\mathrm{L},j}= 1-2\rho_j$. Therefore, the transfer matrix of the normal components of the Poynting vector for each interface $j=1, 2, \dots, k+1$ is
\begin{equation}
\mathbf{N}_j=
\frac{1}{\tau_{{\mathrm R},j}}
\begin{pmatrix}
1 & -\rho_{{\mathrm L},j} \\
\rho_{{\mathrm R},j} & \delta_j
\end{pmatrix}
=\frac{1}{1-\rho_j}
\begin{pmatrix}
1 & -\rho_j \\
\rho_j & 1-2\rho_j
\end{pmatrix}
\label{eq23}
\end{equation}
and thus the resulting transfer matrix $\mathbf{N}$ of the normal components of the Poynting vector with the resulting reflectances and transmittances is given by the product of the sub-matrices in the correct order
\begin{equation}
\mathbf{N}=\mathbf{N}_1 \mathbf{N}_2\dots \mathbf{N}_{k+1}=
\frac{1}{\tau_{\mathrm R}}
\begin{pmatrix}
1 & -\rho_{\mathrm L} \\
\rho_{\mathrm R} & \tau_{\mathrm R}\tau_{\mathrm L}-\rho_{\mathrm R}\rho_{\mathrm L}\
\end{pmatrix}.
\label{eq24}
\end{equation}
Hence, for the resulting reflectance from the left (in this case the same as from the right), we get the expression from the elements of the matrix $\mathbf{N}$
\begin{equation}
\rho_{\mathrm{R}}(n_1,\dots , n_k)=-\dfrac{N_{12}}{N_{11}}=\dfrac{N_{21}}{N_{11}},
\label{eq25}
\end{equation}
which need not be explicitly stated.

For our purpose, we understand $\rho_{\mathrm{R}}$ as a~function of the refractive indices of the layers with given refractive indices of the surrounding media for which we are looking for the minimum. Some numerical minimization methods can be used to find the minimum reflectance of layers with incoherent light, but there is also an analytical solution. It is sufficient to set the derivatives of the function $\rho_{\mathrm{R}}(n_1,\dots , n_k)$ with respect to all variables equal to zero if $n_0$ and $n_s$ are fixed. From this requirement we obtain simple recurrent relations
\begin{equation}
n_j^2=n_{j-1}n_{j+1},\quad j=1, 2, \dots , k
\label{eq26}
\end{equation}
with obvious boundary conditions for $n_0$ and $n_{k+1}=n_s$.

The solution of the equations (\ref{eq26}) can be found in the form
\begin{equation}
n_j=n_0^{1-\frac{j}{k+1}}n_s^{\frac{j}{k+1}},
\label{eq27}
\end{equation}
which again indicates the validity of the relation (\ref{eq21}) given for the coherent situation, which can be considered as a~consequence of the mirror symmetry of the planar system of layers. Note that we can easily generalize the result (\ref{eq27}) for oblique incidence and s or p polarization when we replace the refractive indices of the media with their admittances (\ref{eq1}).

We can also find explicit expressions for the (minimum) size of the reflectance of a~layer system and incoherent light
\begin{equation}
\begin{split}\label{eq28}
&\rho_{\mathrm {min}}=\dfrac{(k+1)(q^{1/(k+1)}-1)^2}{(k+1)q^{2/(k+1)}-2(k-1)q^{1/(k+1)}+k+1},\\
&q=\dfrac{n_0}{n_s}\ \mathrm{or\ \ }q=\dfrac{n_s}{n_0}
\end{split}
\end{equation}
with the same result and expected limit equal to zero for the number of layers $k$ going to infinity.

\section{Results of numerical experiments}
The figures~1--6 present the results of numerical calculations of the exact designed maximally flat antireflection systems for three selected numbers of layers, as described in the theoretical part of the paper. As expected from the requirement of impedance matching for coherent and incoherent electromagnetic fields, the refractive index profiles converge to stepwise smooth transition of their values from the superstrate to the substrate or vice versa as the number of layers increases. However, each of the two cases discussed in detail provides this transition with a~different curve, but always such that the symmetry equation (\ref{eq21}) is satisfied. It can also be seen that the band of low reflectance broadens as the number of layers increases, and in the case of the design for incoherent light the number of local alternating minima and maxima of reflectance increases (values below $10^{-15}$ are not shown in the plots). The attempt to further reduce the reflectivity over a~wider wavelength region of the systems thus designed by varying their thicknesses does not lead to the goal, since the solutions already obtained represent local extremes in themselves.

\begin{figure}[H]\label{fig1}
\begin{center}
\fbox{
\includegraphics[width=7.0cm]{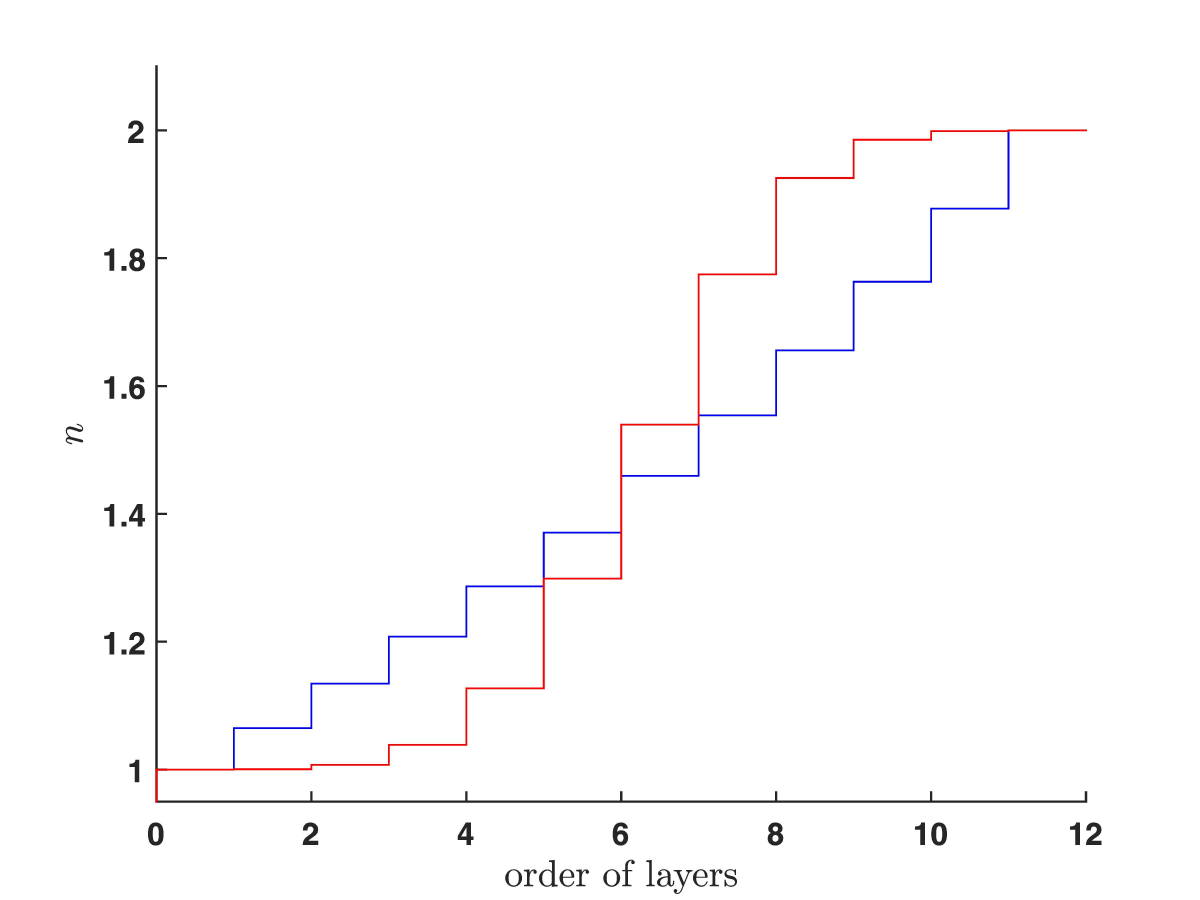}
}
\caption{
Refractive index profiles of 10 dielectric layers realizing maximally flat antireflection for coherent light (red line) compared to layers designed for incoherent light (blue line), substrate refractive index $n_s=2$, superstrate refractive index $n_0=1$.
}
\end{center}
\end{figure}

\begin{figure}[H]\label{fig2}
\begin{center}
\fbox{
\includegraphics[width=7.0cm]{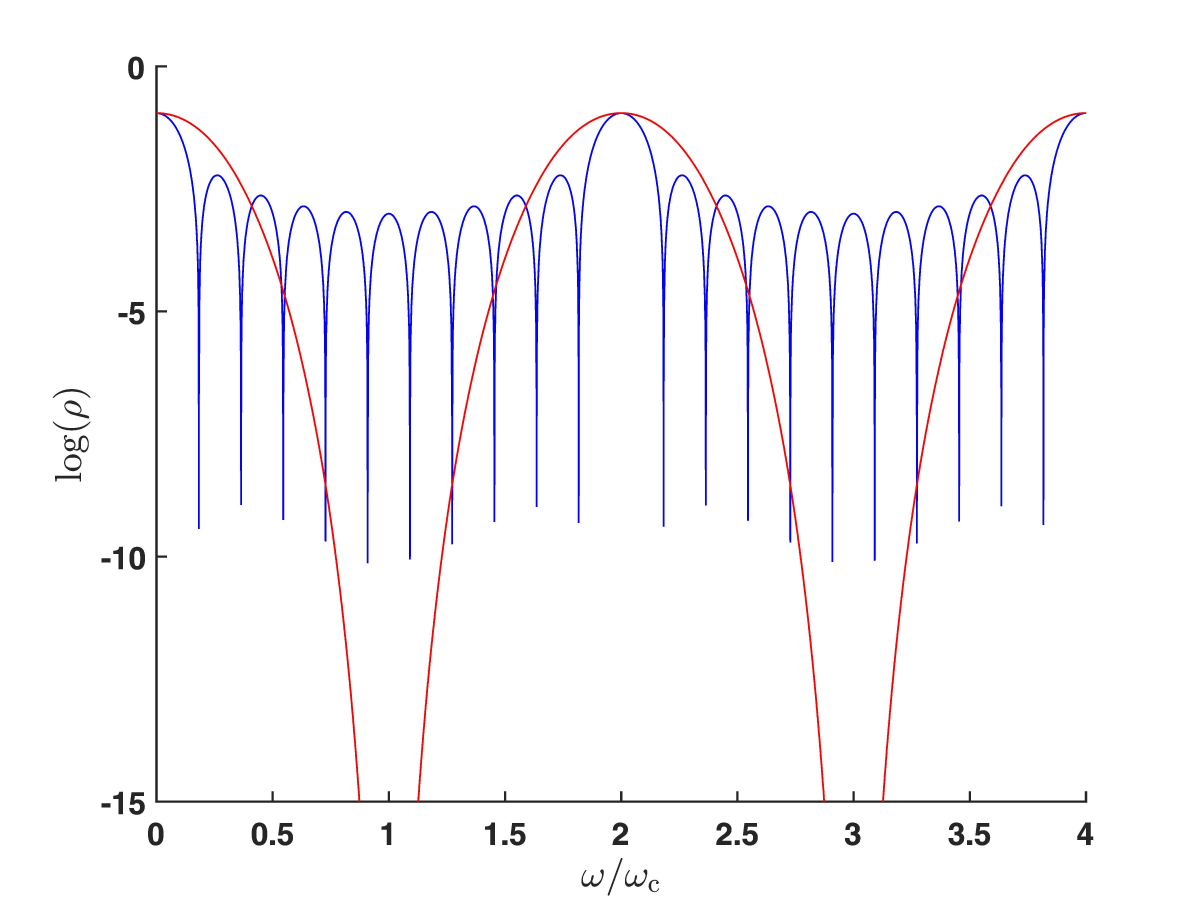}
}
\caption{
Dependence of the decimal logarithm of reflectance $\log(\rho)$ on the relative frequency $\omega/\omega_c$ of the system of 10 quarter-wave layers for the refractive index profiles shown in fig.~1, the red line refers to the refractive indices of the layer system designed for coherent light, the blue line is for the refractive indices of layers designed for incoherent light.
}
\end{center}
\end{figure}

\begin{figure}[H]\label{fig3}
\begin{center}%
\fbox{
\includegraphics[width=7.0cm]{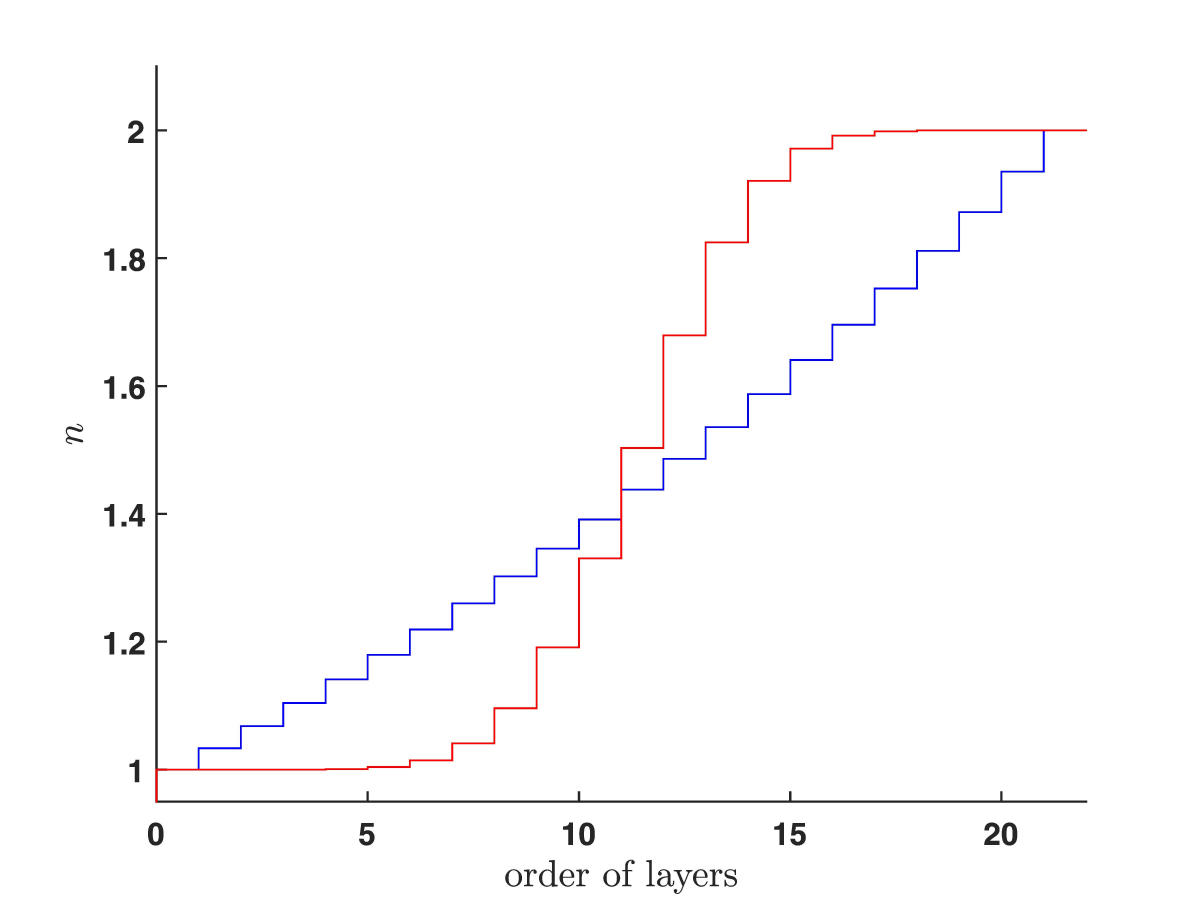}
}
\caption{
Refractive index profiles of 20 dielectric layers realizing maximally flat antireflection for coherent light (red line) compared to layers designed for incoherent light (blue line), substrate refractive index $n_s=2$, superstrate refractive index $n_0=1$.
}
\end{center}
\end{figure}

\begin{figure}[H]\label{fig4}
\begin{center}
\fbox{
\includegraphics[width=7.0cm]{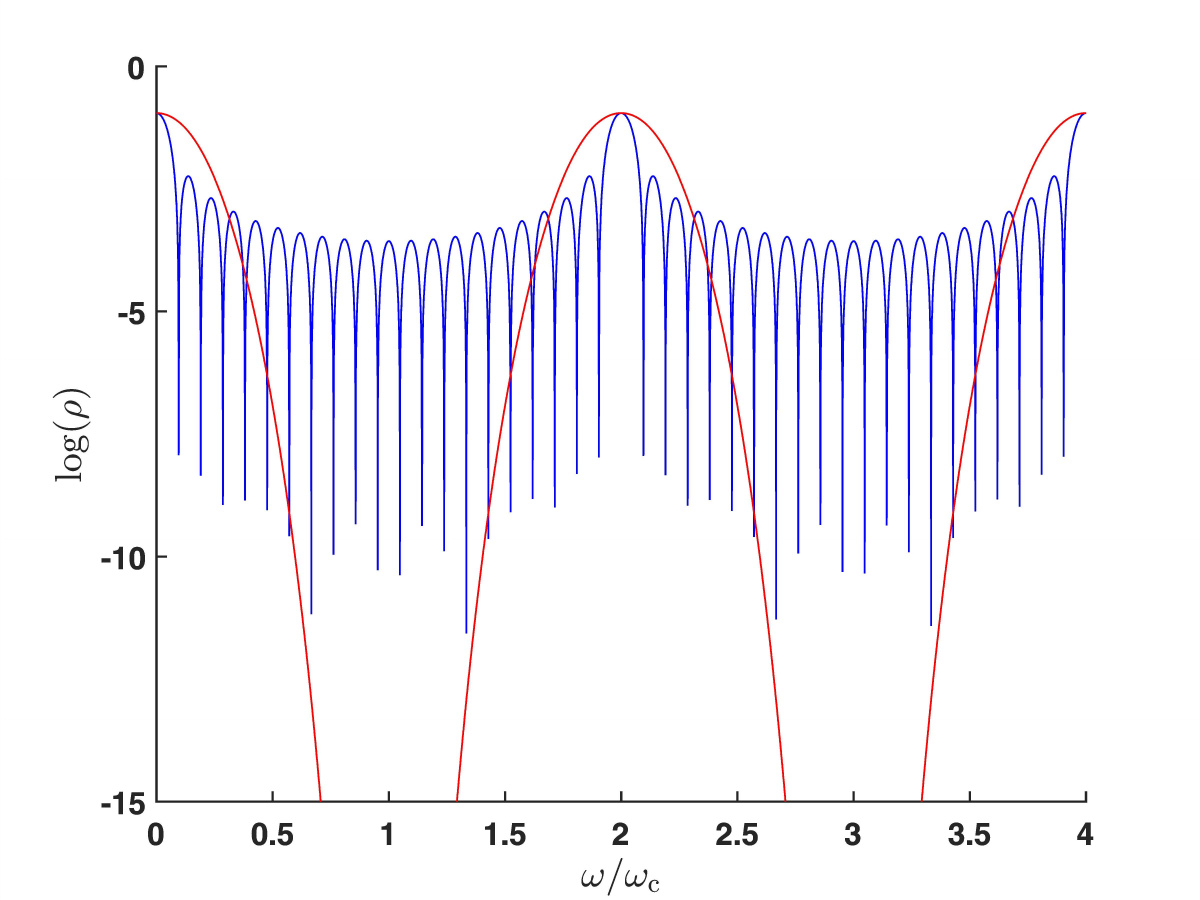}
}
\caption{
Dependence of the decimal logarithm of reflectance $\log(\rho)$ on the relative frequency $\omega/\omega_c$ of the system of 20 quarter-wave layers for the refractive index profiles shown in fig.~3, the red line refers to the refractive indices of the layer system designed for coherent light, the blue line is for the refractive indices of layers designed for incoherent light.
}
\end{center}
\end{figure}

 \begin{figure}[H]\label{fig5}
\begin{center}
\fbox{
\includegraphics[width=7.0cm]{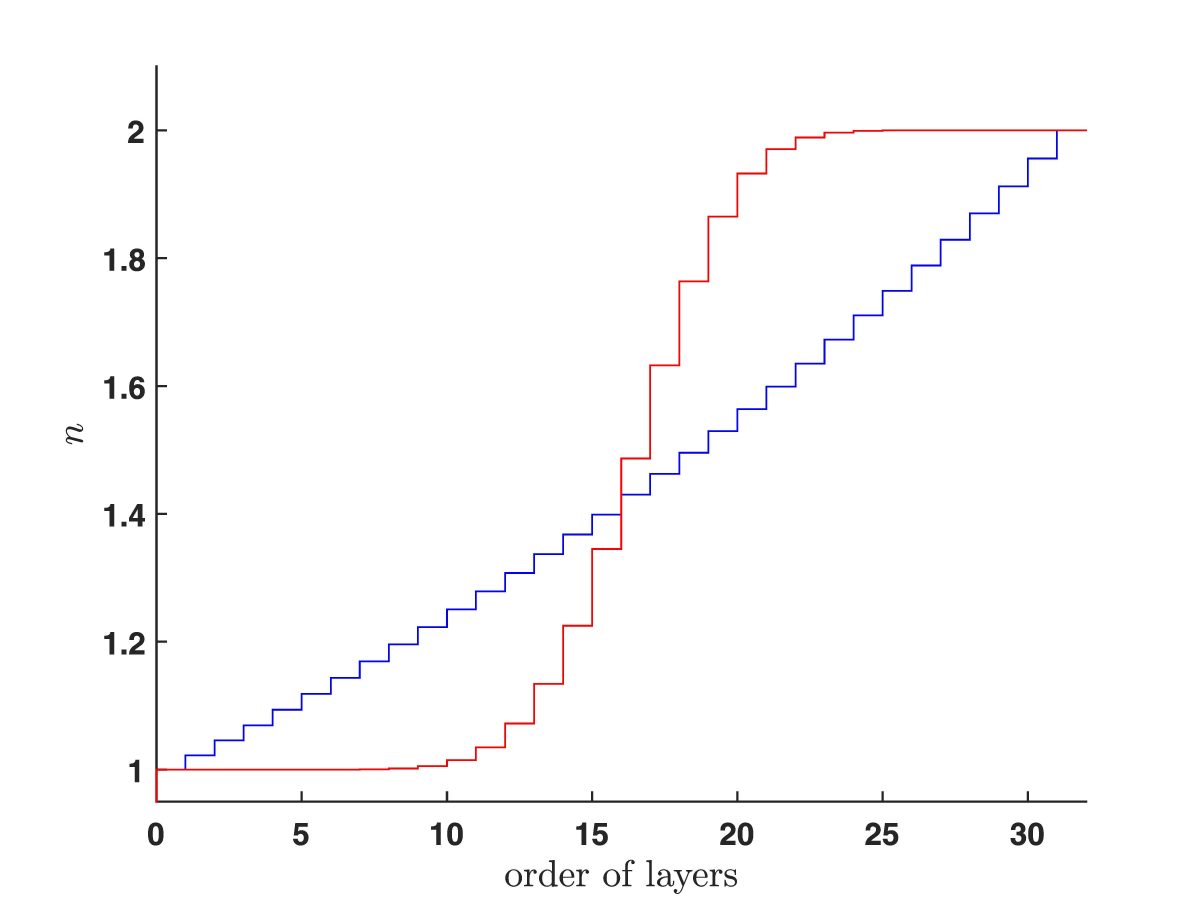}
}
\caption{
Refractive index profiles of 30 dielectric layers realizing maximally flat antireflection for coherent light (red line) compared to layers designed for incoherent light (blue line), substrate refractive index $n_s=2$, superstrate refractive index $n_0=1$.
}
\end{center}
\end{figure}

\begin{figure}[H]\label{fig6}
\begin{center}
\fbox{
\includegraphics[width=7.0cm]{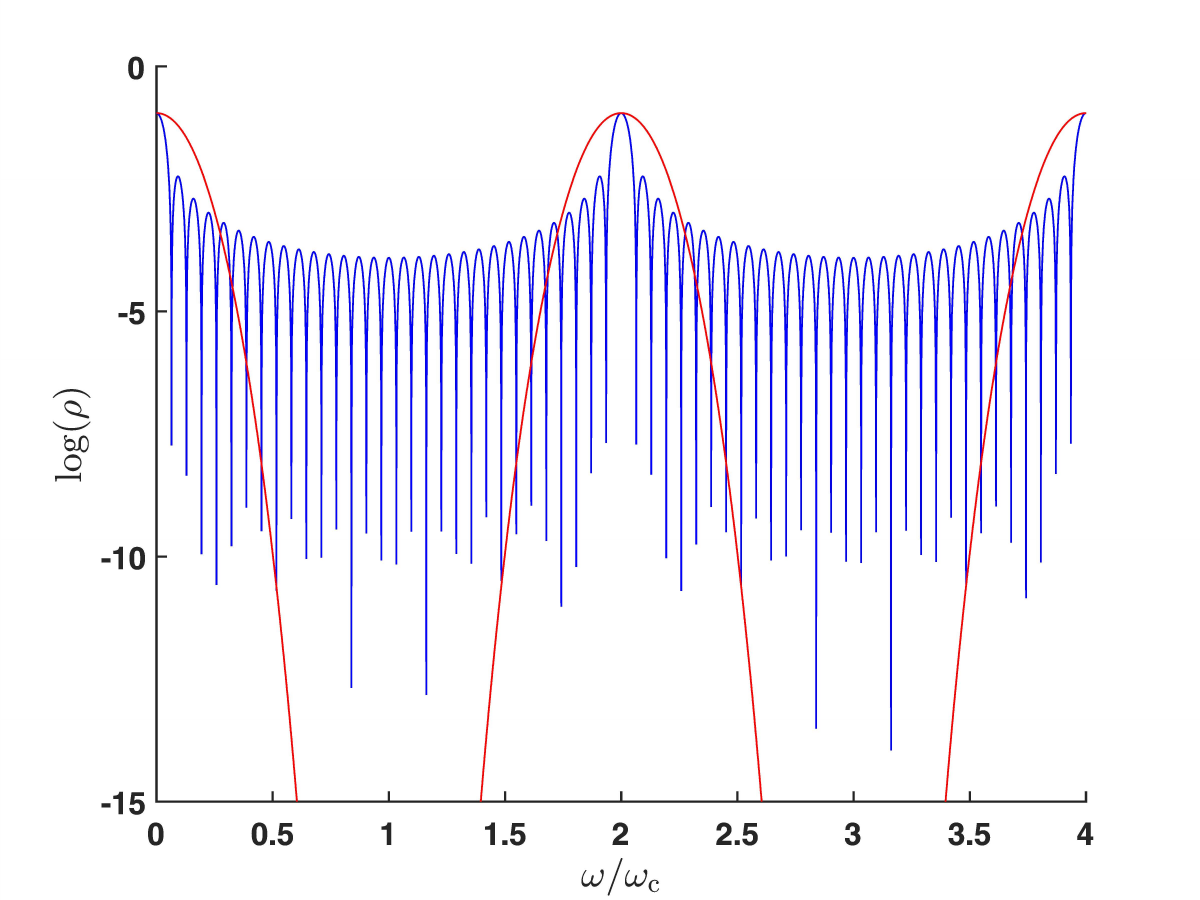}
}
\caption{
Dependence of the decimal logarithm of reflectance $\log(\rho)$ on the relative frequency $\omega/\omega_c$ of the system of 30 quarter-wave layers for the refractive index profiles shown in fig.~5, the red line refers to the refractive indices of the layer system designed for coherent light, the blue line is for the refractive indices of layers designed for incoherent light.
}
\end{center}
\end{figure}

 \begin{figure}[H]\label{fig7}
\begin{center}
\fbox{
\includegraphics[width=7.0cm]{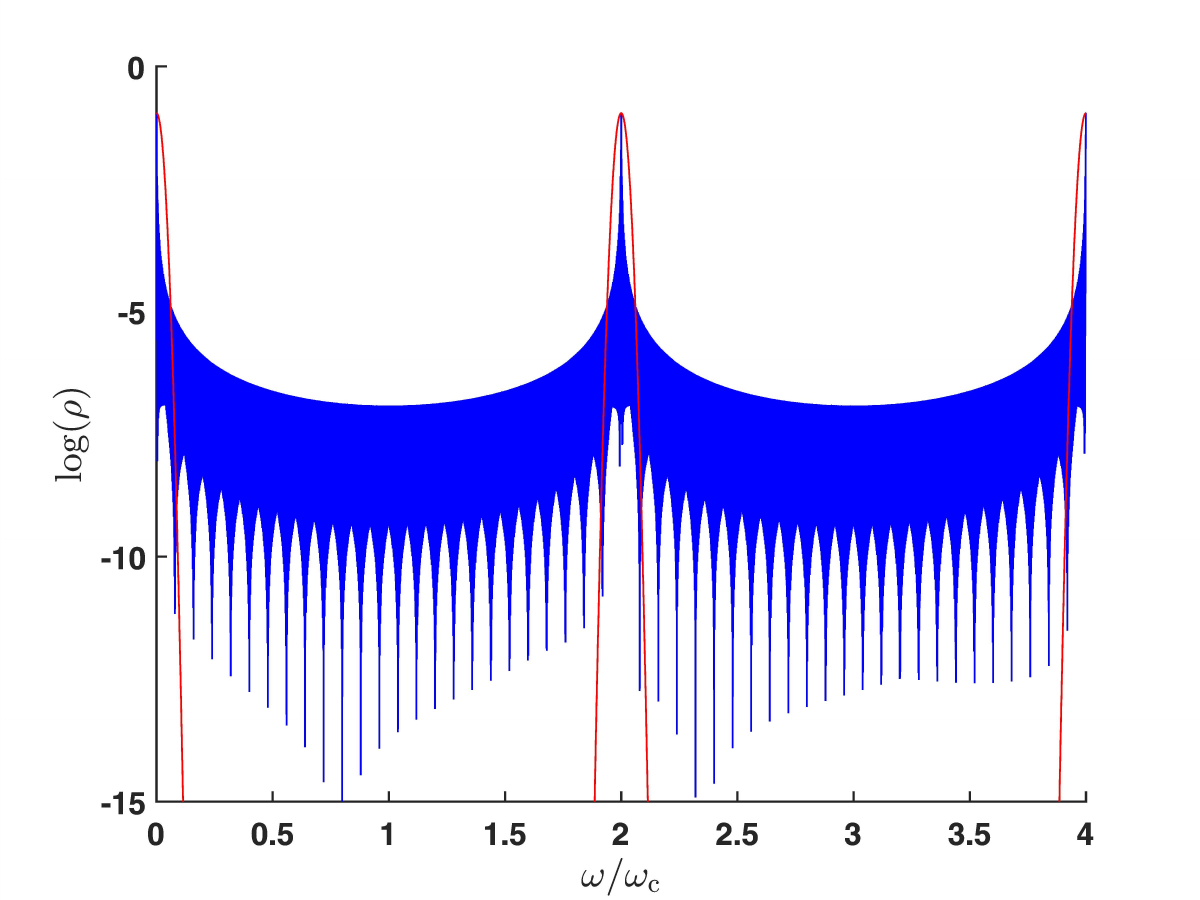}
}
\caption{
Dependence of the decimal logarithm of reflectance $\log(\rho)$ on the relative frequency $\omega/\omega_c$ of the system of 1000 quarter-wave layers designed for maximally flat antireflection, the red line refers to the refractive indices of the layer system designed for coherent light, the blue line refers to the layer system designed for incoherent light.
}
\end{center}
\end{figure}

For illustration, fig.~7 shows the reflectance versus relative frequency $\omega/\omega_c$ of a~system of 1000 quarter-wave layers designed for maximally flat antireflection in the coherent and incoherent cases, the refractive indices of the superstrate and substrate are as in the previous figures. The low reflectance of the order of $10^{-6}$ for incoherent light can be determined from the easily calculated refractive indices (see equation \ref{eq27}), but this is not the case for coherent light due to the large demands on computer time.

\section{Discussion}
The equation (\ref{eq27}) for the refractive indices of the layers in the case of incoherent light satisfies the condition (\ref{eq21}), originally derived for the coherent case. The profile of the refractive indices of the antireflection layers in figs.~1, 3, 5 suggest that it might be possible to match the desired refractive indices with a~suitable curve that would resemble a~Gaussian error function. Indeed, if we assume the sizes of the refractive indices of the layers to be of the form
\begin{equation}\label{eq29}
\begin{split}
&n_j=n_0^{1-\alpha^{(k)}_j}n_s^{\alpha^{(k)}_j},\ j=0,\dots, k+1,\\
&\alpha^{(k)}_0=0,\ \alpha^{(k)}_{k+1}=1.
\end{split}
\end{equation}
For example, from the known refractive indices $n_j$ we get the requirement for the size of $\alpha^{(k)}_j$ 
\begin{equation}
\alpha^{(k)}_j=\dfrac{\log(n_j/n_0)}{\log(n_{k+1}/n_0)},\ j=0,\dots, k+1.
\label{eq30}
\end{equation}
For the incoherent case we obviously have $\alpha^{(k)}_j=j/(k+1)$ from (\ref{eq27}).

The auxiliary coefficients $\alpha^{(k)}_j$ entering the expression for the refractive indices of the layers using the Gaussian error function can be assumed in the approximate form
\begin{equation}
\alpha^{(k)}_j=\dfrac{1}{2} \left[1+\mathrm{erf}\left(b_k(j-\dfrac{k+1}{2})\right)\right],\ j=1,\dots, k.
\label{eq31}
\end{equation}
Since the function $\mathrm{erf}(x)$ is odd, such a~dependence satisfies the condition (\ref{eq21}), now written as $\alpha^{(k)}_j+\alpha^{(k)}_{k+1-j}=1$. However, the coefficients $b_k$ have to be found numerically so that the fitted refractive indices are as close as possible to the solutions of equations (\ref{eq16}), which can be done numerically by the least squares method. The refractive indices of the layers obtained by such a~matching procedure differ from the exact values only at the fourth decimal place, as shown in fig.~8, which would hardly be observable in a~graphical representation of the refractive index profile.

\begin{figure}[H]\label{fig8} 
\begin{center}
\fbox{
\includegraphics[width=7.0cm]{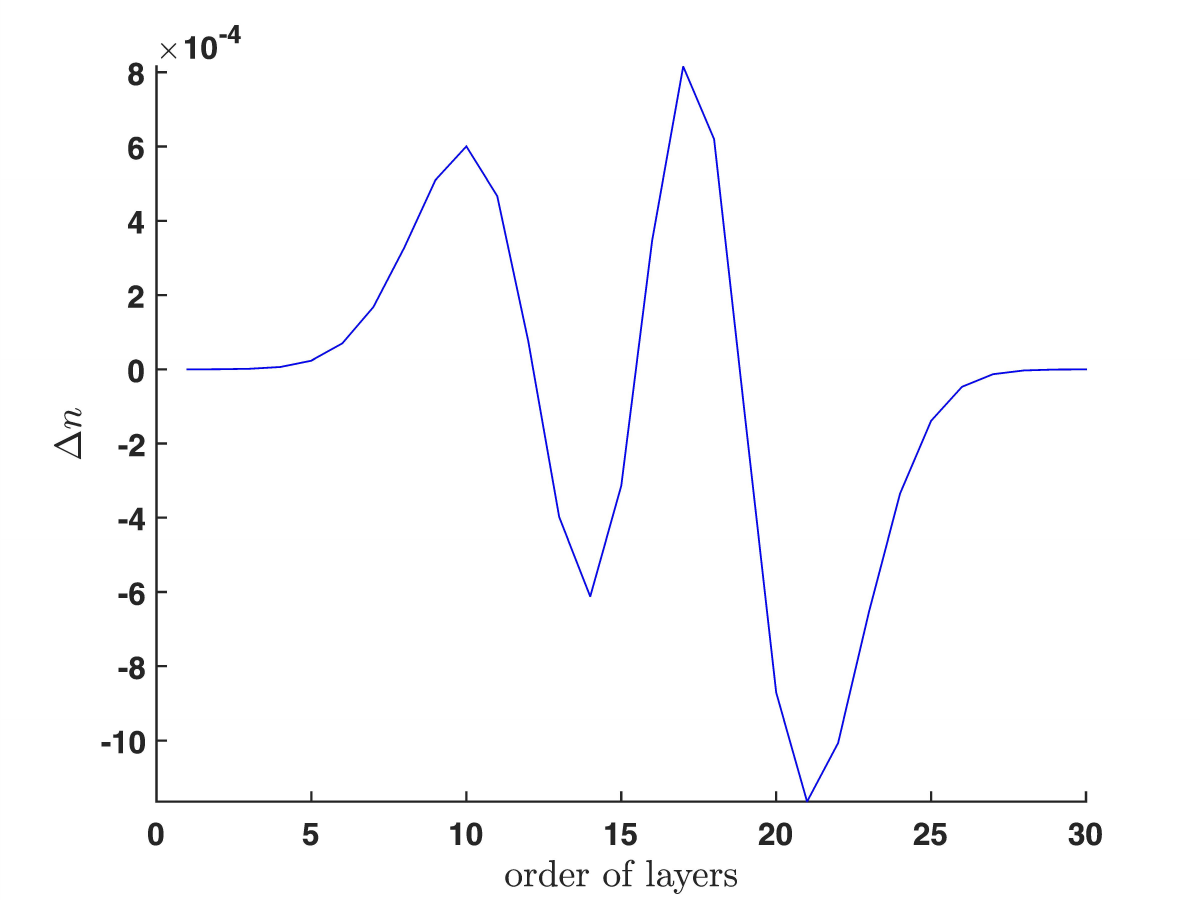}
}
\caption{
The difference $\Delta n$ in the refractive indices of the 30 layers obtained by approximation using the Gaussian error function (\ref{eq31}) and the exact refractive indices calculated by solving the equations (\ref{eq16}) depending on the layer order. The coefficient $b_{30}=0.257959458606224$ entering (\ref{eq31}) was obtained numerically, the refractive indices of the surrounding environments are $n_0=1$, $n_s=2$.
}
\end{center}
\end{figure}

\begin{figure}[H]\label{fig9}
\begin{center}
\fbox{
\includegraphics[width=7.0cm]{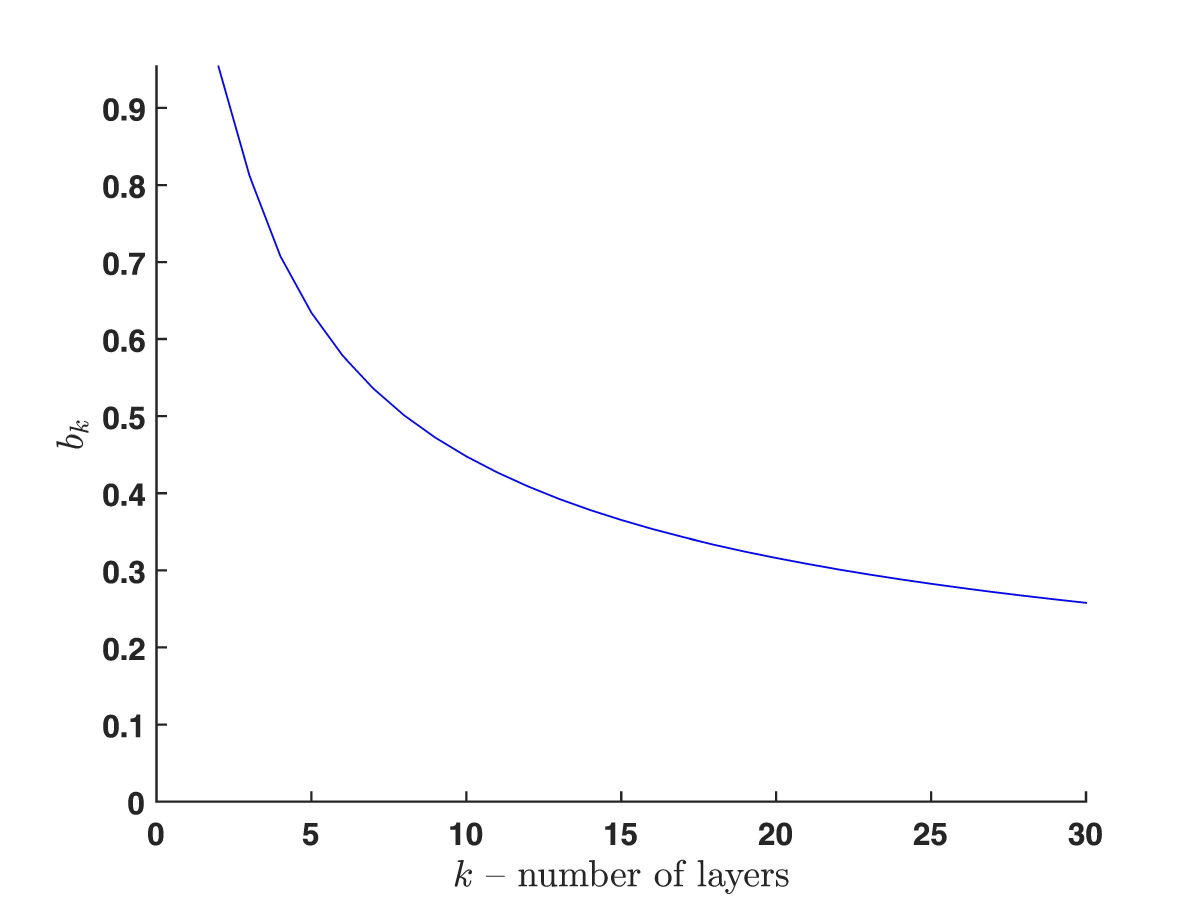}
}
\caption{
The size of the optimized coefficients $b_k$ entering the approximation expression (\ref{eq31}) depending on the number of layers, the refractive indices of the environment surrounding the layers are $n_0=1$, $n_s=2$.
}
\end{center}
\end{figure}

However, the coefficients $b_k$ calculated in this way depend on the number of layers, as shown in fig.~9. We can also assume their dependence on the refractive indices of the surrounding media, but we would welcome this dependence to be as negligible as possible, which would allow us to easily make antireflection designs close to the theoretical best.
 
We can also ask how small differences in approximate refractive indices of the order of $10^{-4}$ from the theoretically accurate values affect the quality of the resulting antireflection. Even if the deviations from the ideal state are small, they are still observable, as shown in fig.~10.

It can be supposed that the stepwise smooth profiles of the refractive indices of the layers according to the relation (\ref{eq31}) will not represent poor antireflection designs, as it is proven by fig.~11 for 100 layers, compared to the ideal refractive indices designed for coherent light, which cannot be easily computed explicitly for 100 layers due to computational complexity.

\begin{figure}[H]\label{fig10}
\begin{center}
\fbox{
\includegraphics[width=7.0cm]{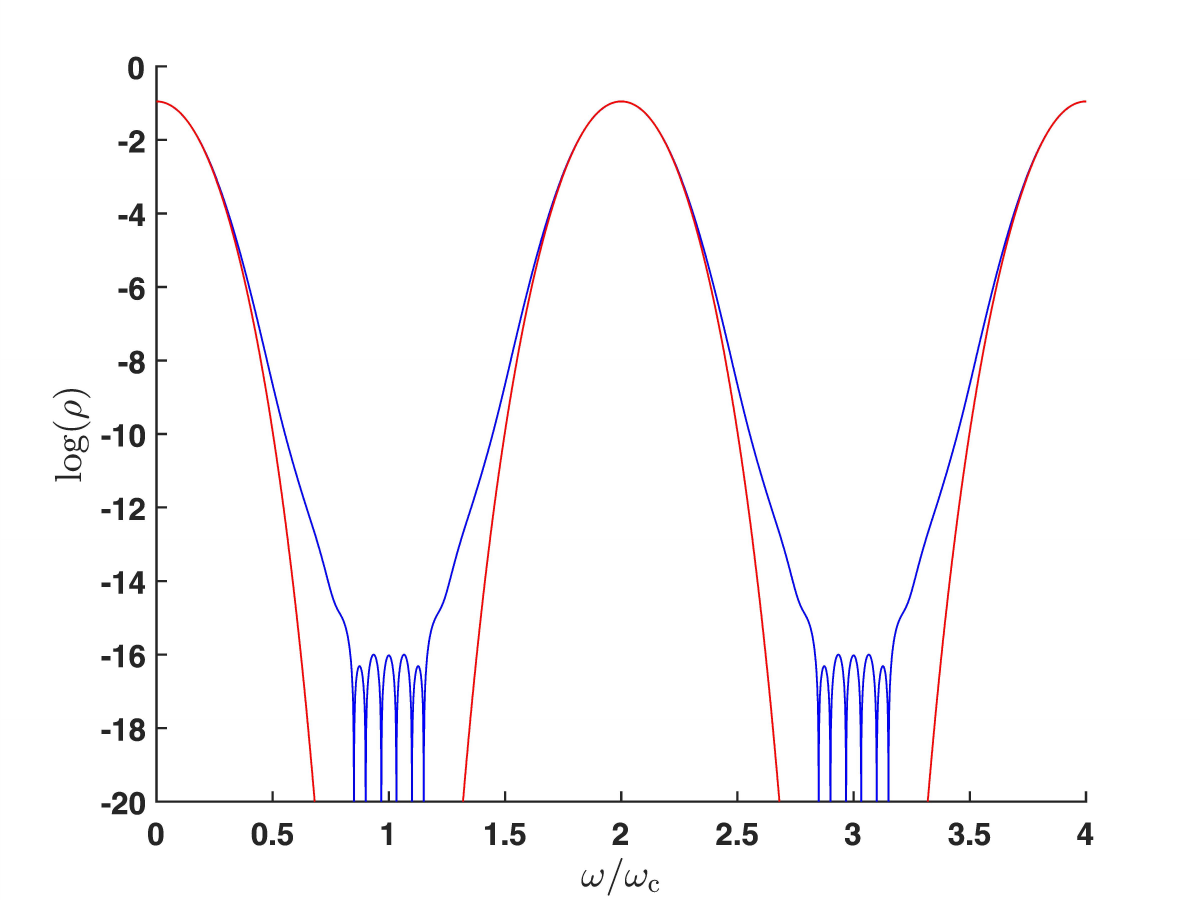}
}
\caption{
Dependence of the decadic logarithm of reflectance $\log(\rho)$ on the relative frequency $\omega/\omega_c$ of a~system of 30 quarter-wave layers designed for maximally flat antireflection in coherent light. The red line refers to the refractive indices of the layers designed according to the exact algorithm, the blue line refers to the refractive indices of the layers designed according to the approximate expression (\ref{eq31}), refractive indices $n_0=1$, $n_s=2$.
}
\end{center}
\end{figure}

\begin{figure}[H]\label{fig11}
\begin{center}
\fbox{
\includegraphics[width=7.0cm]{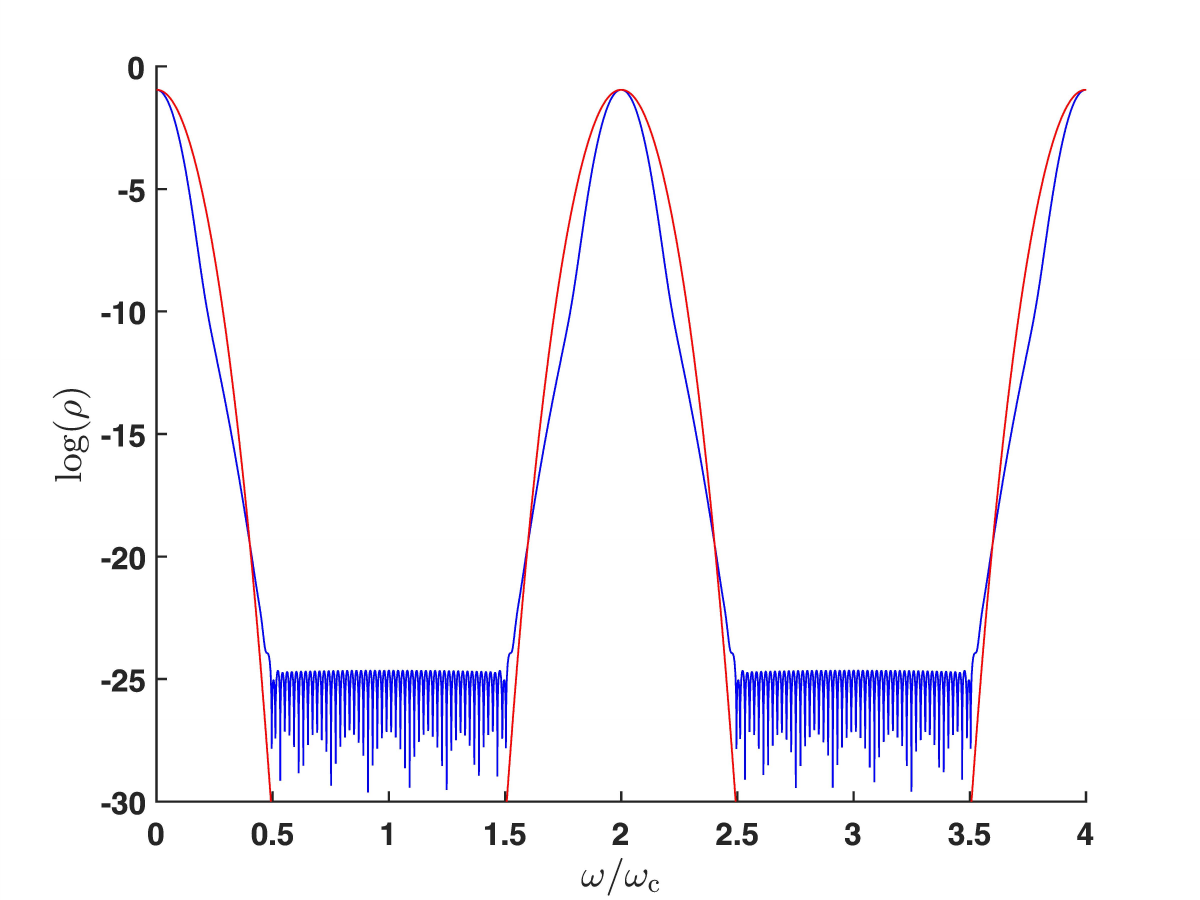}
}
\caption{
Dependence of the decadic logarithm of the reflectance $\log(\rho)$ on the relative frequency $\omega/\omega_c$ of a~system of 100 quarter-wave layers designed according to the relation (\ref{eq31}) with parameter $b_{100}=0.1$ (blue line). For comparison, the red line refers to the refractive indices of the layers designed according to the exact algorithm, $n_0=1$, $n_s=2$.
}
\end{center}
\end{figure}

However, the relation (\ref{eq31}) can be used to design an antireflection with a~different refractive index of the substrate with a~suitable choice of the parameter $b_k$. For example, fig.~12 shows the result of the calculation for 30 layers and $n_0=1$, $n_s=4$ with the choice of $b_{30}=0.258$. Thus, the expression (\ref{eq31}) allows the designer to play with different antireflection constructions that are more or less close to the exact solution.

%kratsi verze
%Perhaps it is interesting to calculate in the coherent case the quantities $P^{(k)}_j=2^k\alpha^{(k)}_j$ and compare them with the Schallenberg's approximation \cite{Schall2008}. If we round the numbers $P^{(k)}_j$ computed numerically according to the exact algorithm to integers, we get the solutions that follow from Schallenberg's procedure. For example, for 4 layers, according to \cite{Schall2008} we have $P^{(4)}=$ 1, 5, 11, 15, while the exact algorithm gives the numbers $P^{(4)}=$ 1.00373255, 5.003732551, 10.996267449, 14.996267449. However, the refractive indices of the layers show small differences. For $n_0=1$, $n_s=2$ we get according to \cite{Schall2008} 1.044273782, 1.241857812, 1.610490332, 1. 915206561, while the exact algorithm yields values of 1.044442656, 1.242058637, 1.610229936, 1.914896897. The relatively small deviations between the exact and Schallenberg's procedures are substantial in order to be able to declare, on the basis of a~comparison of the reflectances around the central wavelength, that maximally flat antireflections for the coherent case provide only solutions of the equations (\ref{eq20}) with the resulting reflectance according to (\ref{eq16}). However, Schallenberg's solution appears to be somewhat better than the approximation using the error function (\ref{eq31}).

%Delsi verze
Perhaps it is interesting to calculate in the coherent case the quantities $P^{(k)}_j=2^k\alpha^{(k)}_j$ and compare them with the Schallenberg's approximation \cite{Schall2008}. Not surprisingly, if we round the numbers $P^{(k)}_j$ computed numerically according to the exact algorithm to integers, we get the solutions that follow from Schallenberg's procedure. For example, for 4 layers, according to \cite{Schall2008} we have $P^{(4)}=$ 1, 5, 11, 15, while the exact algorithm gives the numbers $P^{(4)}=$ 1.003732550742073, 5.003732550742090, 10.996267449257910, 14.996267449257919. However, the refractive indices of the layers show small differences. For $n_0=1$, $n_s=2$ we get according to \cite{Schall2008} 1.044273782427414, 1.241857812073484, 1.610490331949254, 1. 91520656139714, while the exact algorithm yields values of 1.044442655609480, 1.242058637263131, 1.610229936009292, 1.914896896692627. The relatively small deviations between the exact and Schallenberg's procedures are substantial in order to be able to declare, on the basis of a~comparison of the reflectances around the central frequency, that maximally flat antireflections for the coherent case provide only solutions of the equations (\ref{eq20}) with the resulting reflectance according to (\ref{eq16}). However, Schallenberg's solution is undoubtebly slightly better than the approximation using the error function (\ref{eq31}).

\begin{figure}[H]\label{fig12}
\begin{center}
\fbox{
\includegraphics[width=7.0cm]{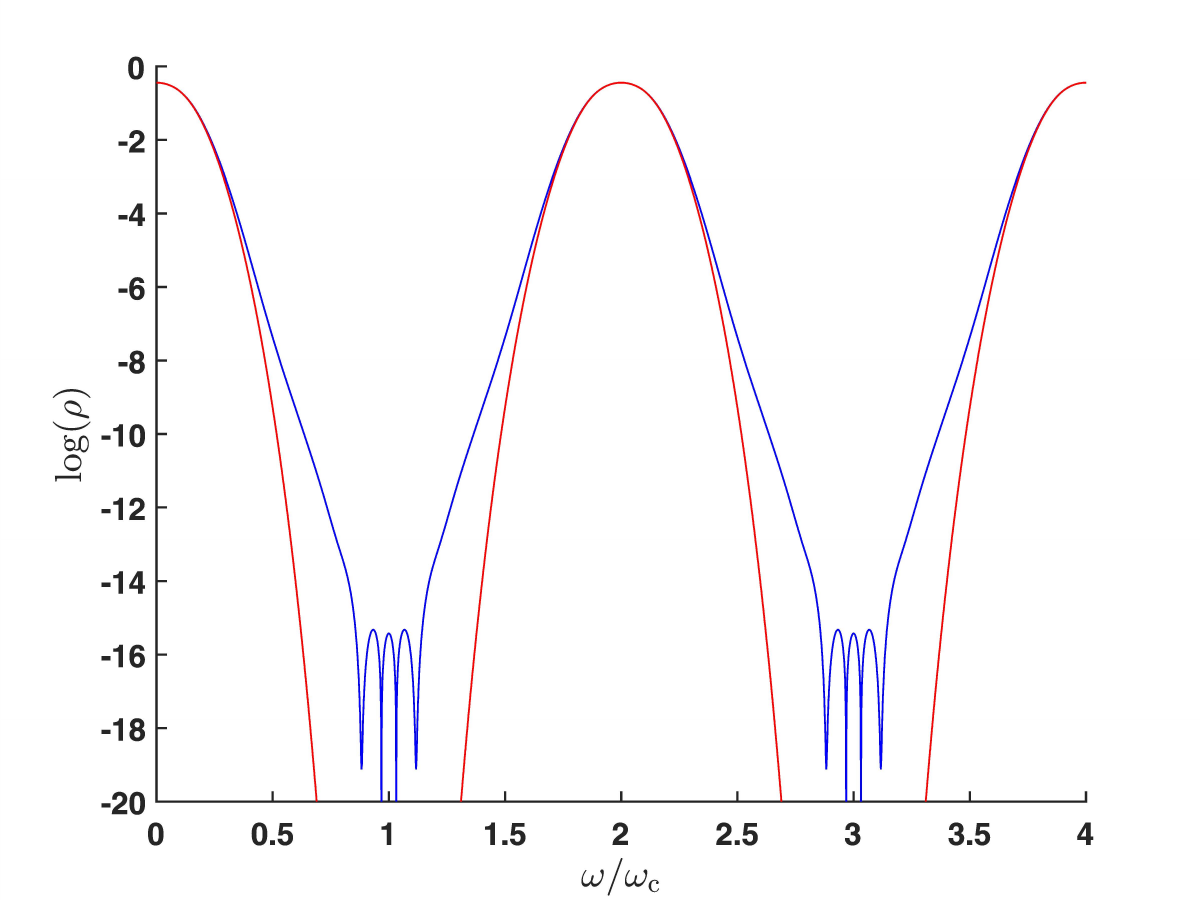}
}
\caption{
Dependence of the decadic logarithm of the reflectivity $\log(\rho)$ on the relative frequency $\omega/\omega_c$ of a~system of 30 quarter-wave layers designed according to the relation (\ref{eq31}) with parameter $b=0.258$ (blue line). For comparison, the red line refers to the refractive indices of the layers designed according to the exact algorithm, $n_0=1$, $n_s=4$.
}
\end{center}
\end{figure}

\section{Conclusions}
In this paper two exact solutions of maximally flat antireflection layer systems are proposed for two ideal cases of coherent (for thin layers) and incoherent (for thick layers) light. The first case is based on the requirement that the maximum number of derivatives of the reflectance of the system of quarter-wave thin layers with respect to the frequency is equal to zero for the central frequency, while the second case is based on the requirement of the minimum reflectance of the system of thick layers. The first case requires finding the zero point of a~system of functions of several variables, which can be done explicitly for systems of at most four layers, while the second case requires finding the minimum of a~function of several variables for which an explicit solution has been found. The graphs then present the obtained profiles of the refractive indices and the corresponding reflectance curves as a~function of the normalized light frequency. There is no speculation about the realization of such profiles, although it cannot be ruled out that the current development of nanotechnology will ultimately be able to achieve flat antireflection to the maximum extent. The possibility of using approximate expressions for layer refractive indices to easily construct quite good antireflection systems is also discussed.

\putacknowledgement

\end{multicols}

% Vlozeni kontaktnich informaci, informaci o vedeckem clanku, anotace v anglictine

%\newpage % pouze pokud nutne

\putcontacts

%\putresearch
\newpage
%\makeannotation 

\end{document}